\documentstyle[11pt,newpasp,twoside,graphicx,epsf]{article}
\def\chandra{{\it Chandra~}}
\def\ergs{erg s$^{-1}$}

\def\edcomment#1{\iffalse\marginpar{\raggedright\sl#1\/}\else\relax\fi}
\reversemarginpar

\begin{document}
\title{ \chandra Observations of the Stellar Populations and Diffuse
Gas in Nearby Galaxies}
 \author{Andreas Zezas, G. Fabbiano, A. Prestwich}
\affil{Harvard-Smithsonian Center for Astrophysics, 60 Garden Street,
Cambridge, MA 02138, USA}
\author{M. Ward}
\affil{University of Leicester, University Road, LE1 7RH, Leicester,
UK}
\author{S. Murray}
\affil{Harvard-Smithsonian Center for Astrophysics}

\begin{abstract}
We present \chandra observations 
 of  two star-forming  galaxies (M~82 and the Antennae) 
and three starburst/AGN composite galaxies
(NGC~1808, NGC~6240, NGC~7331).
In both star-forming galaxies we detect a large number of sources
with diverse properties. Some of them can be identified as X-ray 
binaries, based on their variability and spectra. However,
there is a significant number of very soft and/or extended sources
which could be supernova remnants. 
 These observations  confirm previous indications
that there is a population of sources with X-ray luminosities much
higher than the Eddington limit for a neutron star, suggesting that
these objects are abundant in star-forming galaxies. 
We find the the X-ray luminosity functions of the discrete sources in
these two galaxies are very similar.
In the case of the composite galaxies we find that the AGN does not
dominate their X-ray emission. A significant 
fraction of the emission from these objects
is extended but there are also X-ray  sources associated with
circumnuclear star-formation.
\end{abstract}
\section{Introduction}
Earlier X-ray imaging observations of normal/star-forming galaxies, showed that their X-ray
emission is very complex, arising
from two different components: (a) discrete sources associated
with supernova remnants (SNRs) and X-ray binaries and (b) diffuse gas often in the form of a
superwind (e.g. Fabbiano 1989; Read, Ponman, \& Strickland 1997). However, the relatively poor spatial resolution of these 
observations hampered any attempt to further study the properties  of each
component in detail. \chandra
is revolutionizing this field by providing X-ray data over a wide
energy band, with an unprecedented spatial resolution
($\sim0.5''$). This spatial resolution is also critical to deconvolve
the starburst and the AGN components in  composite galaxies. 
We present the first results from  \chandra
 observations of two well studied star-forming galaxies (M82 and the
Antennae) and three composite  starburst/AGN galaxies (NGC1808,
NGC6240 and NGC7331). 
\section{The star-forming galaxies: M82 and The Antennae}
\subsection{M82}
 The prototypical star-forming galaxy M~82,  was observed on four
occasions with the \chandra Advanced CCD Imaging Spectrometer (ACIS-I),
with exposures ranging between 5 and 15~ks.
 It was also observed twice with the  \chandra High Resolution Camera
(HRC).
\begin{figure}
\plotone{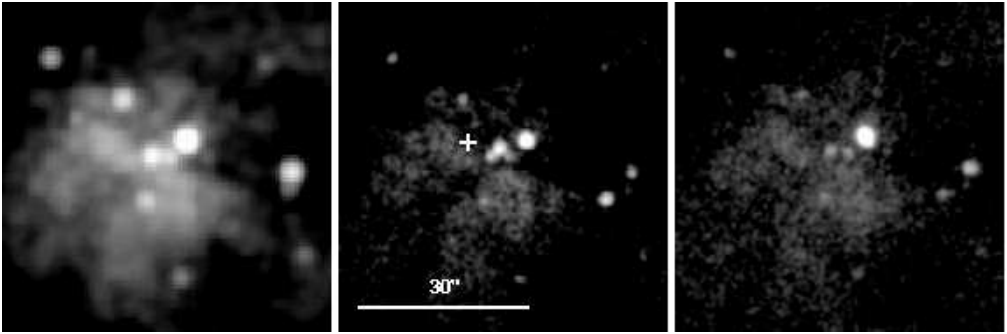}
\caption{The central (0.9$'$ x 0.9$'$) region of M82 as observed in
September 1999 (left), October 1999 (middle) and January 2000
(right).  The  first two images share the same color-map, whereas in the third the
color-map  is stretched in order to show the fainter sources. The cross
in the second image marks the dynamical center of the galaxy. In this
and the following images North is up and East is to the left. }
\end{figure}
The \chandra observations show that there are 24 discrete sources down
to a detection limit of $\sim10^{37}$ \ergs. These sources comprise 10\%
of the total soft (0.3 - 2.5 keV) X-ray emission and 50\% of the total
hard (2.5 - 10.0 keV) X-ray emission of the galaxy. Twelve  sources are found
to vary by factors of 20 - 700\% on timescales of 1 to 6 months, with
the brightest of them reaching a peak luminosity of $\sim10^{41}$
\ergs. The \chandra HRC observations of this extremely luminous source
have been reported by Kaaret et al. (2000) and Matsumoto et al.
(2000) and the first results from one ACIS-I observation are presented
by Griffiths et al. (2000).  Figure~1 shows three observations of the central kpc of M82
taken a few months apart. The first observation was performed
with ACIS-I  and the other two with HRC. 
 The variability of some of these sources suggests that they most
probably are associated with X-ray binaries (although, 
compact SNRs cannot be entirely dismissed).  Three more sources are
coincident with SNRs detected in deep radio observations.  
We constructed two cumulative X-ray luminosity function (XLF) of the
sources in M82: one from the peak luminosity of each source and one
from the longest observation. When fitted with a power-law they have comparable slopes
($\alpha=-0.40\pm0.04$ and $\alpha=-0.45\pm0.06$ respectively). 

\subsection{The Antennae (NGC~4038/9)}
 The Antennae are the prototypical example of merging galaxies. They
were observed with ACIS-S for 72ks.
\begin{figure}
\plotone{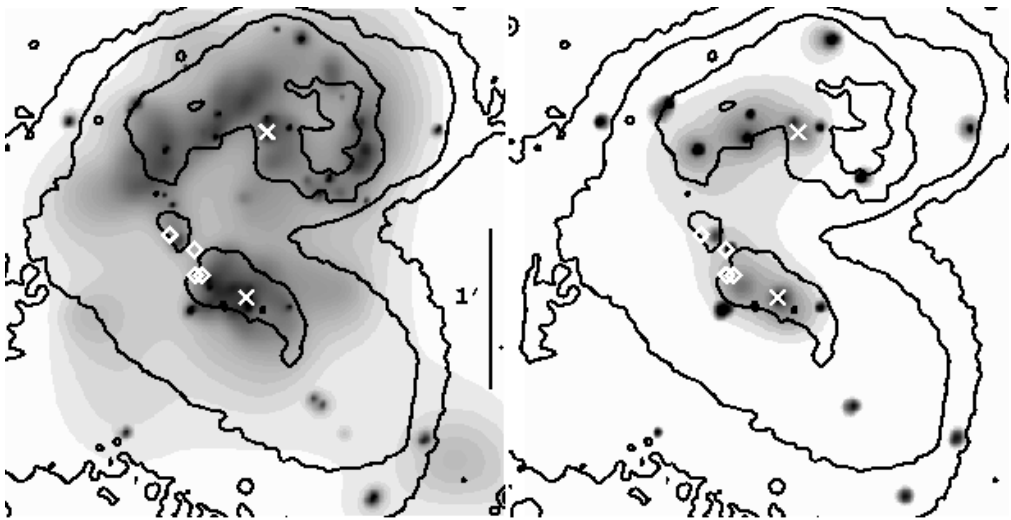}
\caption{ An adaptively smoothed soft (0.3-2.0~keV; left) and medium
(2.0-4.0~keV; right) band X-ray image  of the Antennae. The contours
show the optical outline of the galaxy. The diamonds and the crosses show the position of the CO
peaks and the two nuclei.}
\end{figure}
We detect a total of 49 sources down to a detection limit of
$\sim10^{38}$ \ergs. The soft (0.3-2.5 keV) X-ray emission is
mostly diffuse (68\% of the total), whereas the hard X-ray emission (2.5-10.0
keV) is dominated by the discrete sources which account for $\sim80$\%
of it.  Figure~2 shows a soft and medium band image of the Antennae,
with overlaid the optical outline of the galaxy. The first results
from this observation have been reported by Fabbiano, Zezas, \& Murray (2001).
 Two sources are found to be variable during this observation and two
more sources are variable in timescales of a few years. The spectra of
the discrete sources are very diverse. There is a marginally
significant  trend for the low
luminosity sources to be softer than higher luminosity sources.  The
co-added spectra of the high luminosity sources ($\rm{L_X}>10^{39}$),
are best fit with a multi-temperature disk black body model together
with a power-law and a thermal plasma model, suggesting that
most of them are associated with X-ray binaries.
 The inner temperature of
the disk is $\sim1.3$ keV, similar to what is found for other very
luminous X-ray sources (eg Makishima et al. 2000). On the other hand
the spectra of the sources with luminosities between $3\times10^{38}$
\ergs and $10^{39}$~\ergs are well fit with a similar model but with a
temperature  of $0.25$ keV for the inner part of the disk, which is
more consistent with stellar size black holes.
The  XLF of the discrete sources in the Antennae galaxies has a slope
of $\alpha=-0.45\pm0.05$, similar to what is found for M82.
Finally, we find a total of 13
X-ray sources with one or more possible bright optical counterparts. All but 
 2 of the optical sources are young  stellar clusters (age $<30$
Myrs). We also find that 22 X-ray sources have one or more radio
counterparts down to a flux level of $\rm{\sim40~\mu Jy}$ at 6~cm. 

\section{The composite galaxies: NGC~1808, NGC~6240 and NGC~7331}
\subsection{NGC~1808}
 NGC~1808 (fig. 3a) is a nearby (10.9~Mpc) composite galaxy. It has a
circumnuclear star-forming ring with a radius of $\sim5''$. 
 In order to disentangle the starburst from the AGN
component  we observed it with the HRC-I for 25.5 ks. We find
a point-like source coincident with the optical nucleus, with an X-ray  
luminosity of $1.2\times10^{39}$ erg/s (0.1-10.0~keV),
contributing only $\sim10$\% of the total X-ray emission of the galaxy.
Apart from the nucleus we detect two more sources which are coincident
with two  HII regions in the circumnuclear starburst. These sources 
 have a luminosity of $1.1\times10^{39}$ erg/s and  are embedded in 
diffuse emission. This diffuse component is most probably due to
 gas heated by the numerous supernovae in the starforming region.  
\begin{figure}
\plottwo{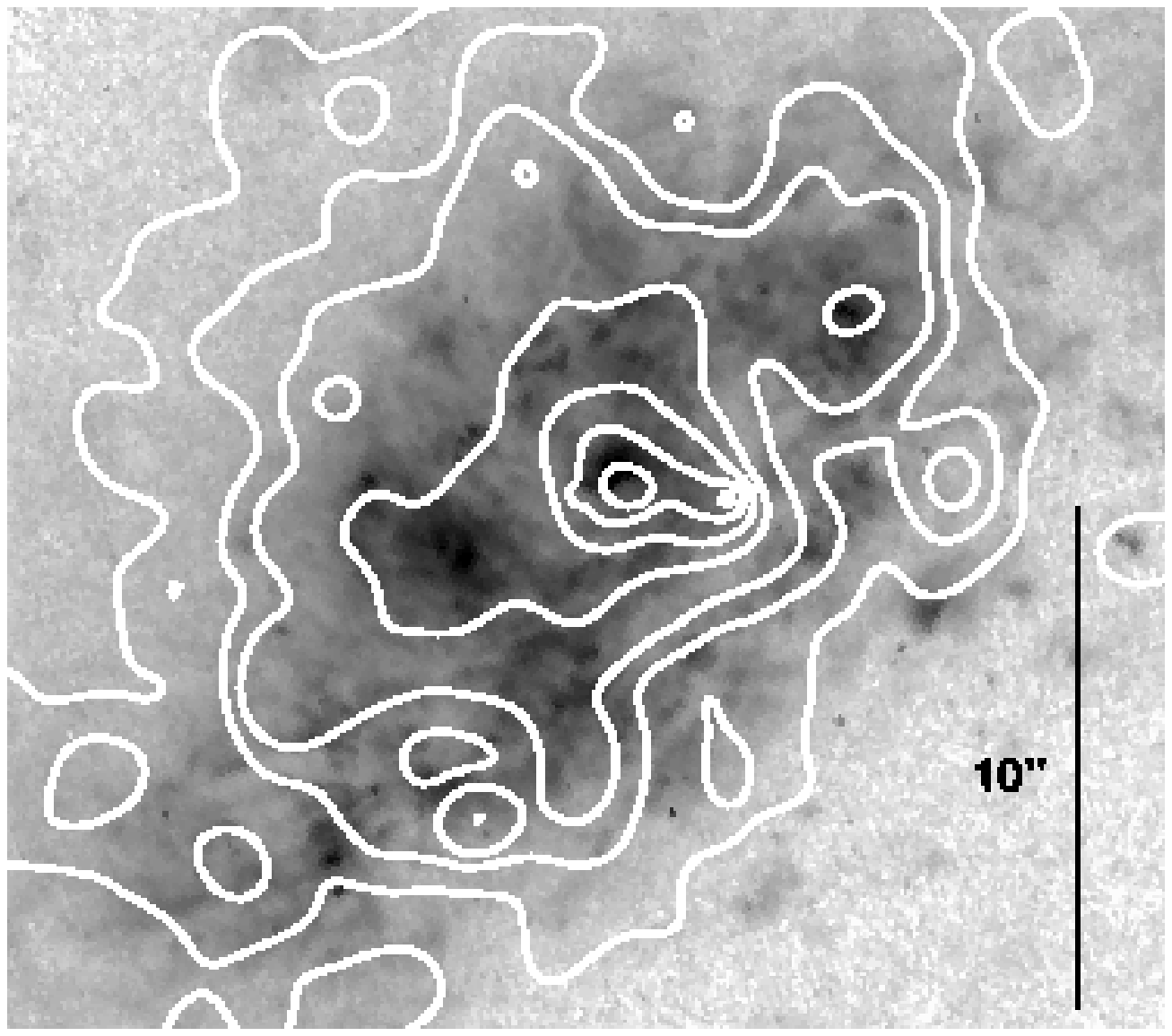}{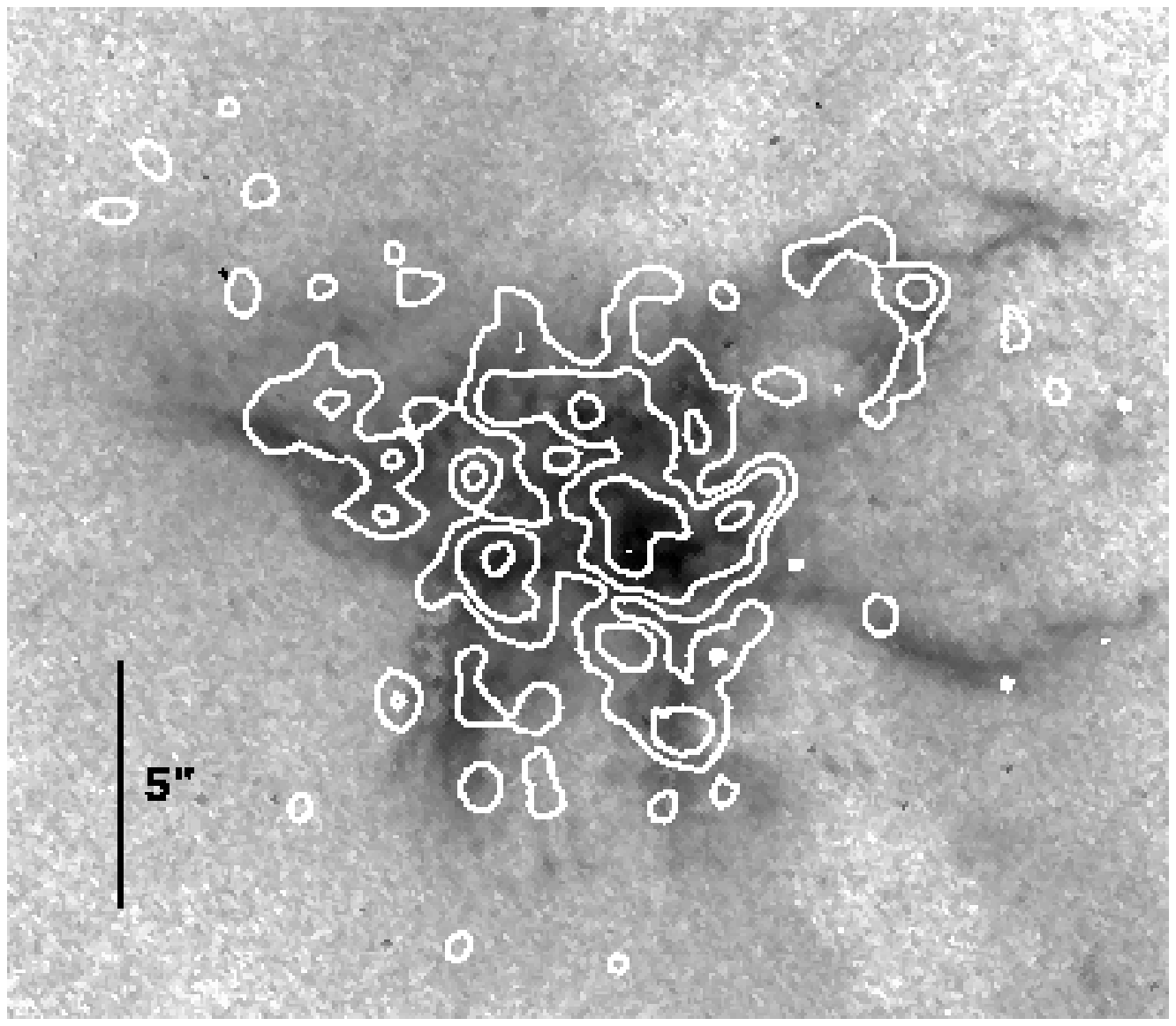}
\caption{HST WFPC2 H$\alpha$ images of NGC~1808 (left) and NGC~6240
(right) with overlaid X-ray contours from the HRC observation.}
\end{figure}
\subsection{NGC~6240}
 NGC~6240 is one of the best studied starburst/AGN composite
galaxies. Previous X-ray observations showed that it hosts a highly
obscured Compton thick AGN (Iwasawa \& Comastri, 1998; Vignati et al.
1999). We observed this galaxy for 8.7 ks with HRC-I. We do not find a
strong nuclear X-ray source; instead the nuclear emission is diffuse
and has a luminosity of $2.4\times10^{41}$ erg/s (for a distance of
100 Mpc), corresponding to 10\% of the total
emission. This  is consistent with this nucleus being Compton
thick, and  suggests that we are viewing either scattered emission
from the AGN or diffuse gas
 associated with the circumnuclear
starburst. Since the HRC does not have any spectral
resolution it is not possible to distinguish between these two
possibilities.
 We do not find many discrete sources, which could be due to the
large distance of NGC~6240 and to the relatively short
exposure. However, we find a striking similarity between the
$\rm{H_{\alpha}}$ and the X-ray morphology (figure ~3b). The most
interesting feature is an arc with a luminosity of $1.7\times10^{41}$
erg/s  in the North-West of the nucleus also seen
 in $\rm{H_{\alpha}}$. This loop could be
 an outflow or a supershell associated with the circumnuclear
starburst. 
\subsection{NGC~7331}
 NGC~7331 is a nearby (14.3 Mpc)  composite galaxy. We observed
it with the ACIS-S camera for 29.5~ks. Figure~4a presents the soft band
(0.3-2.0 keV) adaptively smoothed image of the galaxy with its optical
outline shown by the  contours.  We detect a total of  35 X-ray
sources down to a luminosity threshold of $\sim5\times10^{37}$~\ergs. 
The XLF of the discrete sources in this galaxy has a slope of
$\alpha=-0.67$. This is steeper than the slope found for the
starforming galaxies and closer to that of early type galaxies,
suggesting that there is a significant contribution from a older
 X-ray source population.

\section{Discussion and Conclusions}
 \chandra allows us to study  the discrete
source populations in galaxies outside the Local Group, in detail  for
the first time. The first
results from these studies show two things: (a) The ultraluminous X-ray
sources ($\rm{L_{X}>10^{39}}$ \ergs) previously found in nearby
galaxies (eg. Fabbiano et al. 1989; Makishima et al. 2000),
are quite common at least among star-forming galaxies; (b) the X-ray
luminosity functions of the discrete sources in star-forming galaxies
are different than those of early type galaxies. 
\begin{figure}
\plottwo{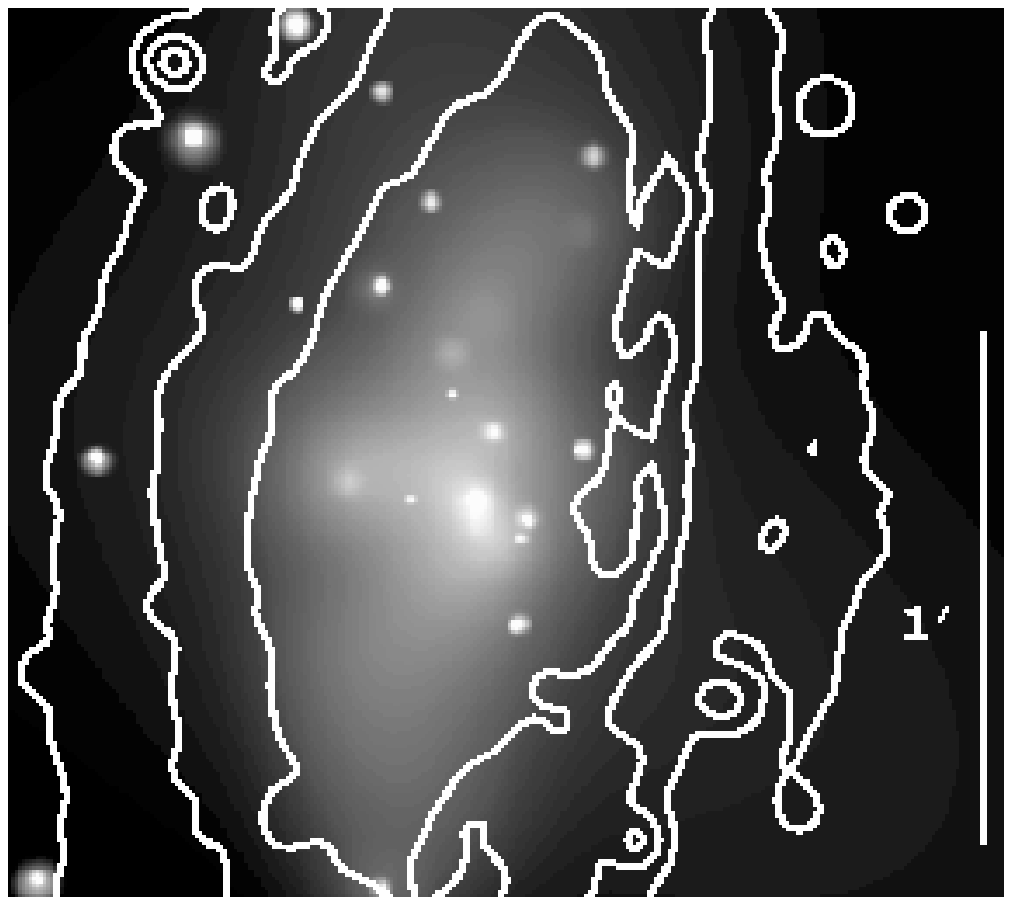}{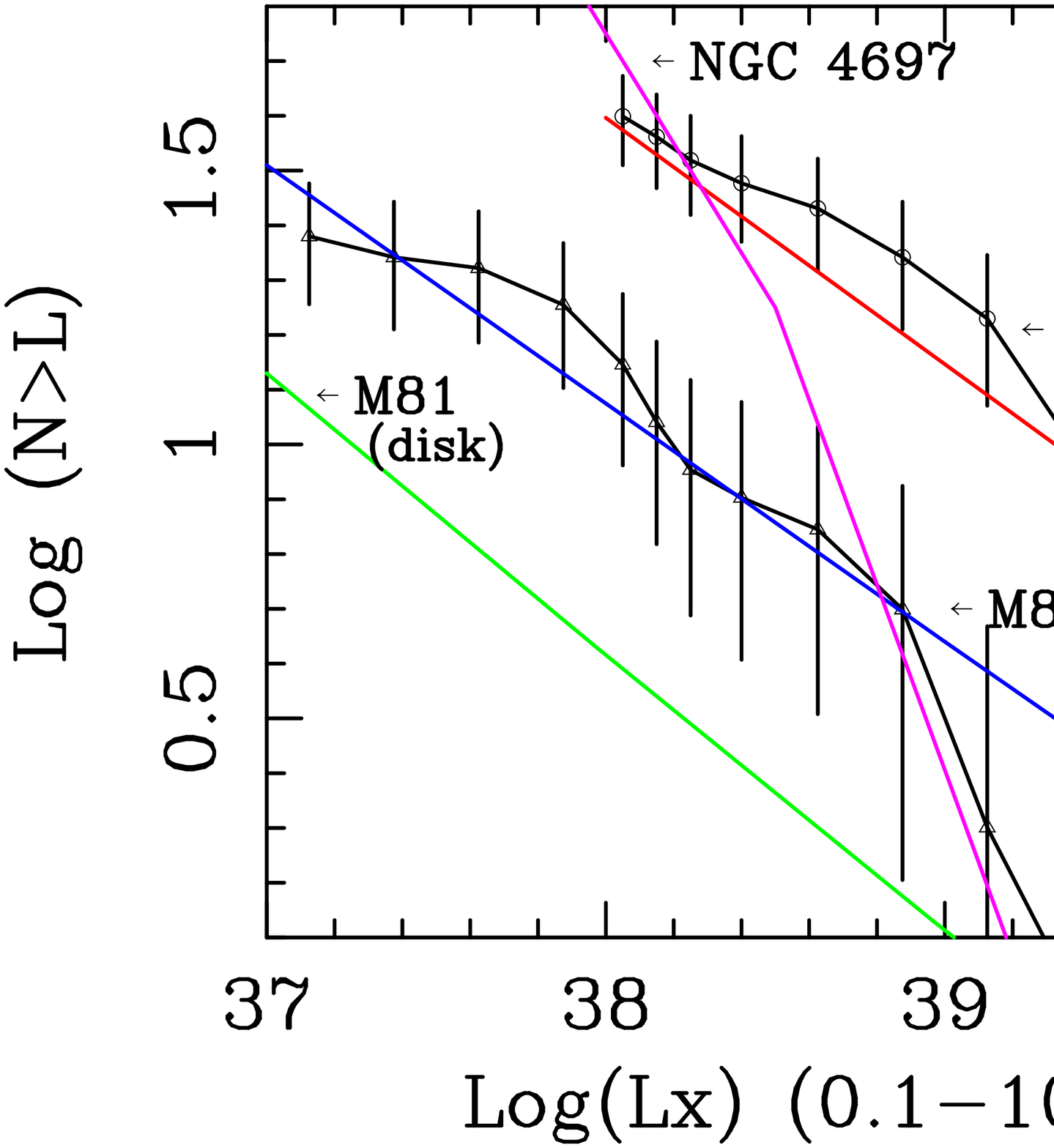}
\caption{Left: An adaptively smoothed soft (0.3-2.0~keV) X-ray image
of the central region of NGC~7331. The contours show the optical
outline of the galaxy. Right: The cumulative XLF of M~82 and the
Antennae shown together with the XLFs of NGC~4697 and M~81.} 
\end{figure}

 In the case of M~82 and the Antennae we detect a total of 3 and 19
ultra-luminous sources (ULXs) respectively. Although some of them could be compact
SNRs (eg. Fabian \& Terlevich, 1996), it is unlikely that most of them are
associated with this class of rare objects. If these
sources are not beamed, the compact object must be a black hole 
in the mass range $10 - 1000$~M\sun. The spectra of similar sources detected in
other galaxies have been interpreted as evidence for accretion onto a
Kerr black hole (eg. Makishima et al. 2001). Suggested models for the
formation of  these intermediate mass black holes,
 include: remnants of Population~III stars (Madau \& Rees, 2001), 
 mergers of smaller black holes (eg. Taniguchi et al.
2000), and mergers of massive stars (Portegies-Zwart, 1999). Another
possibility is that the intrinsic luminosity of these sources is
mildly beamed (King et al. 2001). In this case the compact
object could be a stellar size black-hole or even a neutron
star. 

  As presented previously, both the XLFs of M~82 and the Antennae have
very similar slopes. This is more clearly seen if Figure~4b which shows
the two luminosity functions as well as the XLFs of the early type
galaxy NGC~4697 (Sarazin, Irwin, \& Bregman 2001) and the nearby spiral
galaxy M~81 (Tennant et al. 2001). In this figure we also see that
the XLFs of the star-forming galaxies are similar to that of M~81, but
very different from that of NGC~4697.  This suggests that 
 X-ray sources associated with a young stellar population
(HMXBs) dominate in  star-forming galaxies. 
Also, we see an excess of ULXs in the more actively star-forming
galaxies (Antennae and M~82) in comparison to the less active M~81.
Although this is very preliminary, it suggests that may be 
 a correlation between the
 presence of ULXs and the level of star-formation activity in a galaxy.

 The observations of the three composite galaxies presented here
clearly show that the AGN does not necessarily dominate the X-ray
emission of the galaxy. This could be due either to an intrinsically
weak AGN (as e.g. in  NGC~7331) or due to large obscuration
(e.g. NGC~6240), as suggested by e.g.  Fabian et al. (1998) and
Levenson et al. (this volume).

\acknowledgments
 We thank Phil Kaaret and Vicky Kalogera for useful discussions. 
 This work has been supported by NASA contracts NAS~8--39073 (CXC)
and NAS8-38248 (HRC) and \chandra grant G01-2116X.

\end{document}